\documentclass{PoS}
\usepackage{cite,epsfig,bm}

\usepackage{array}
\newcolumntype{T}{>{\ttfamily} c}
\newcolumntype{M}{>{$\displaystyle} c <{$}}

\title{
\vspace*{-1.2cm}
\begin{minipage}{\textwidth}
{\normalfont\small LTH 1174, Nikhef 2018-042, DESY 18-151
\hspace{\fill} August 2018}\\
\end{minipage}\\[10pt]
Anomalous dimensions and splitting functions beyond
the next-to-next-to-leading order}

\ShortTitle{Anomalous dimensions and splitting functions beyond NNLO}

\author{\speaker{A. Vogt}\\
        \mbox{Department of Mathematical Sciences, University of Liverpool,
        Liverpool L69 3BX, UK}\\
        E-mail: \email{Andreas.Vogt@liverpool.ac.uk}
        }

\author{F. Herzog \phantom {g} \\
       \mbox{Nikhef Theory Group, Science Park 105, 1098 XG Amsterdam,
       The Netherlands}\\
       E-mail: \email{fherzog@nikhef.nl}
       }

\author{S. Moch \phantom {g} \\
       II.~Institut f\"ur Theoretische Physik, Universit\"at Hamburg,
       D-22761 Hamburg, Germany\\
       E-mail: \email{sven-olaf.moch@desy.de}
       }

\author{B. Ruijl \phantom {g} \\
       Institute for Theoretical Physics, ETH Z\"urich, 8093 Z\"urich,
       Switzerland\\
       E-mail: \email{bruijl@phys.ethz.ch}
       }

\author{T. Ueda \phantom {g} \\
       Department of Materials and Life Science, Seikei University,
       Musashino, Tokyo 180-8633, Japan\\
       E-mail: \email{tueda@st.seikei.ac.jp}
       }

\author{J.A.M. Vermaseren \phantom {g} \\
       \mbox{Nikhef Theory Group, Science Park 105, 1098 XG Amsterdam,
       The Netherlands}\\
       E-mail: \email{t68@nikhef.nl}
       \\ \\ \\ }

\abstract{ 
We report on recent progress on the splitting functions for the 
evolution of parton distributions and related quantities, the 
(lightlike) cusp anomalous dimensions, in perturbative QCD.
New~results are presented for the four-loop 
(next-to-next-to-next-to-leading order, N$^3$LO) contributions
to the flavour-singlet splitting functions and the gluon cusp
anomalous dimension. 
We present first results, the moments $N=2$ and $N=3$, 
for the five-loop (N$^4$LO) non-singlet splitting functions.}  

\FullConference{Loops and Legs in Quantum Field Theory (LL2018)\\
		29 April 2018 - 04 May 2018, St. Goar, Germany}

\newcommand{\beq}{\begin{equation}}
\newcommand{\eeq}{\end{equation}}
\newcommand{\bea}{\begin{eqnarray}}
\newcommand{\eea}{\end{eqnarray}}
\newcommand{\nn}{\nonumber}

\def\frct#1#2{\mbox{\small{$\displaystyle\frac{#1}{#2}$}}}

\newcommand{\Qeq}{\raisebox{-0.15mm}{$\:\!\stackrel{Q_{\phantom{i\!\!\!\!}}}{=}\:\:$}}
\newcommand{\ra}{\rightarrow}

\newcommand{\ar}{a_{\sf s}}
\newcommand{\al}{\alpha_{\sf s}}
\def\als(#1){{\alpha_{\sf s}^{\:\!#1}}}
\def\ars(#1){{a_{\sf s}^{\:\!#1}}}

\newcommand{\ep}{\varepsilon}
\newcommand{\MSb}{$\overline{\mbox{MS}}$}

\def\nc{{n_c}}
\def\ncs{{n_{c}^{\,2}}}

\def\ca{{C^{}_A}}
\def\cas{{C^{\,2}_A}}
\def\cat{{C^{\,3}_A}}
\def\caf{{C^{\,4}_A}}
\def\cf{{C^{}_F}}
\def\cfs{{C^{\, 2}_F}}
\def\cft{{C^{\, 3}_F}}
\def\cff{{C^{\, 4}_F}}
\def\nf{{n^{}_{\! f}}}
\def\nfz{{n^{\,0}_{\! f}}}

\def\nfs{{n^{\,2}_{\! f}}}
\def\nft{{n^{\,3}_{\! f}}}

\def\xm1{{(1 \! - \! x)}}

\def\zr#1{{\zeta_{#1}^{}}}

\definecolor{Blue}{rgb}{0.00,0.00,1.00}

\begin{document}

\section{Introduction}

\noindent
Up to power corrections, observables in {\it ep} and {\it pp} hard 
scattering can be schematically expressed~as
\bea
\label{OepOpp}
  O^{\,ep} \;=\; f_{i}^{} \,\otimes\, c_i^{\,\rm o} 
\; , \quad
  O^{\,pp} \;=\; f_{i}^{} \,\otimes\, f_{k}^{} 
                 \,\otimes\, c_{ik}^{\,\rm o}
\eea
in terms of the respective partonic cross sections (coefficient 
functions) $c^{\,\rm o}$ and the universal parton distribution 
functions (PDFs) $f_{i\,}^{}(x,\mu^2)$ of the proton at a 
scale $\mu$ of the order of a physical scale.
The dependence of the PDFs on the momentum fraction $x$ is not 
calculable in perturbative QCD; their scale dependence is given 
by the renormalization-group evolution~equations
\beq
\label{Evol}
 \frct{\partial}{\partial \ln \mu^2} \: f_i^{}(x,\mu^2) \: =\: 
 \int_x^1 \frct{dy}{y} \: P^{}_{ik}\big(y,\al(\mu^2)\big) \, 
 f_k^{}\Big(\,\frct{x}{y}\,,\mu^2 \:\!\!\Big) 
\; .
\eeq

\vspace*{1mm}
The splitting functions $P^{}_{ik\,}$, which are closely related to 
the anomalous dimensions of \mbox{twist-2} operators in the light-cone 
operator-product expansion (OPE), and the coefficient functions in 
eq.~(\ref{OepOpp}) can be expanded in powers of the strong coupling
$\ar \equiv \al(\mu^2)/(4\pi)$,
\bea
\label{Pexp}
 P \, & \:=\: & \quad
    \ars()\, P^{\,(0)}
       +\, \ars(2)\, P^{\,(1)}
       +\, \ars(3)\, P^{\,(2)}
     \,+\, \ars(4)\, P^{\,(3)}
       +\: \ldots
\;\;\; , \\
\label{Cexp}
 c^{\,\rm o}_{a} & \:=\: & \ar^{\:n_{\rm o}}
   \big[ \, c_{\rm o}^{\,(0)}
   \;+\: \ars()\, c_{\rm o}^{\,(1)}
   \;+\: {\ars(2)\, c_{\rm o}^{\,(2)}} \,
   \:+\: {\ars(3)\, c_{\rm o}^{\,(3)}}
   \:+\: \ldots \,\big]
\; .
\eea
Together, the first three terms of eqs.~(\ref{Pexp}) and (\ref{Cexp}) 
provide the NNLO approximation for the observables (\ref{OepOpp}).
This is now the standard accuracy of perturbative QCD for many hard 
processes; see refs.~\cite{P2,DP2} for the corresponding 
helicity-averaged and helicity-dependent splitting functions.
N$^3$LO corrections have been obtained for inclusive lepton-hadron 
deep-inelastic scattering (DIS) \cite{MVV610+}, Higgs production in 
proton-proton collisions \cite{Higgs1,Higgs2}, and jet production
in DIS \cite{N3LOjet}.
N$^4$LO results for inclusive DIS have been reported 
in refs.~\cite{SumRule} (sum rules) and ref.~\cite{avLL16} 
(low Mellin-$N$ moments).
  
\vspace*{1mm}
Using basic symmetries, the system (\ref{Evol}) can be decomposed 
into $2\:\!\nf-\!1$ scalar `non-singlet' equations and a 
$2 \!\times\! 2$~flavour-singlet system.
The former includes $2 (\nf-\!1)$ flavour asymmetries of 
quark-antiquark sums and differences, $q_i^{} \pm \bar{q}_i^{}$, 
and the total valence distribution,
\beq
\label{qns}
  q_{{\rm ns},ik}^{\,\pm} \; = \; 
  q_i^{} \pm \bar{q}_i^{} \,-\, (q_k^{} \pm \bar{q}_k^{})
\:\: , \quad
  q^{}_{\rm v} \; = \;
  {\textstyle \sum_{\,r\,=1}^{\,\nf}} \, (q_r^{} - \bar{q}_r^{})
\; .
\eeq
The singlet PDFs and their evolution are given by
\beq
\label{qPsg}
  q^{\,}_{\,\sf s} \; = \;
  {\sum_{r\:\!=\:\!1}^{\,\nf}} \, (q_r^{} + \bar{q}_r^{})
\; , \quad
  \frac{d}{d \ln\mu^2} \,
  \bigg( \begin{array}{c} \! q^{}_{\:\!\sf s} \! \\[-0.5mm] \! g\!
         \end{array} \bigg)
  \; = \; \bigg( \! \begin{array}{cc}
       \, P_{\rm qq} \, & P_{\rm qg} \, \\[-0.5mm]
       \, P_{\rm gq} \, & P_{\rm gg} \, \end{array} \! \bigg) \otimes
  \bigg( \begin{array}{c} \!q^{}_{\,\sf s}\! \\[-0.5mm] \! g\!
         \end{array} \bigg)
\; ,
\eeq
where $g(x,\mu^2)$ denotes the gluon distribution.
$P_{\rm qq}$ differs from the splitting function $P_{\rm ns}^{\,+}$ 
for the combinations $q_{{\rm ns},ik}^{\,+}$ in Eq.~(\ref{qns}) by a
pure singlet contribution $P_{\rm ps}^{}$ which is suppressed at 
large $x$.
In~this limit, the splitting functions $P_{\rm qq}$ and $P_{\rm gg}$ 
in the standard \MSb\ scheme are of the form
\beq
\label{xto1}
  P_{\rm kk}^{\,(n-1)}(x) \;=\;
        \frac{x\,A_{n,\rm k}}{(1-x)_+}
  \,+\, B_{n,\rm k} \, \delta \xm1
  \,+\, C_{n,\rm k} \, \ln \xm1
  \,+\, D_{n,\rm k} 
  \,+\, \xm1\,\mbox{-terms} 
\; ,
\eeq
where $A_{n,\rm q}$ and $A_{n,\rm g}$ are the (light-like) $n$-loop 
quark and gluon cusp anomalous dimensions \cite{Korch89}. 
These and the `virtual anomalous dimensions' $B_{n,\rm k}$ are 
relevant well beyond the context of Eq.~(\ref{Evol}).
 
\vspace*{1mm}
In this contribution we briefly report on recent N$^3$LO (4-loop) 
results for the singlet splitting functions in eq.~(\ref{qPsg}), 
including the gluon cusp anomalous dimension $A_{4,\rm g}$ 
\cite{MRUVV2}, 
and on the first N$^4$LO \mbox{(5-loop)} calculations of the 
non-singlet splitting functions $P_{\rm ns}^{\,\pm}$.
For the (more advanced) status of the 4-loop non-singlet splitting
functions the reader is referred to refs.~\cite{DRUVV,MRUVV1,avRadC17}.

\section{Low-$\bm N$ results for the N$\bm ^3$LO singlet splitting 
functions}

\noindent
The results for $N=2$ and $N=4$ have been reported, in numerical form 
for $\nf = 4$ flavours in~QCD, at the previous Loops \& Legs Workshop 
\cite{avLL16}.
In the meantime, the computations of four-loop DIS with {\sc Forcer} 
\cite{Forcer}, which are conceptually straightforward extensions of
the three-loop calculations in refs.~\cite{3loopMoms}, 
have been extended to $N=6$ for 
$P_{\,\rm qg}^{\,(3)}$ and $P_{\,\rm gg}^{\,(3)}$ 
and to $N=8$ for 
$P_{\,\rm qq}^{\,(3)}$ and~$P_{\,\rm gq}^{\,(3)}$.
The resulting perturbative expansions of $P_{ik}^{}(N,\nf=4)$ 
are approximately given by 
\bea
\label{Pqqnf4}
  P_{\,\rm qq}^{}(2,4) &\,=\,& - 0.28294\, \als() \left(
  1 \,+\, 0.6219\, \als()
    \,+\, 0.1461\, \als(2)
    \,+\, 0.3622\, \als(3) 
    \,+\, \ldots \, \right)
\; , \nn \\
  P_{\,\rm qq}^{}(4,4) &\,=\,& - 0.55527\, \als() \left(
  1 \,+\, 0.6803\, \als()
    \,+\, 0.4278\, \als(2)
    \,+\, 0.3459 \, \als(3) 
    \,+\, \ldots \, \right)
\; , \nn \\
  P_{\,\rm qq}^{}(6,4) &\,=\,& - 0.71645\, \als() \left(
  1 \,+\, 0.6489\, \als()
    \,+\, 0.4264\, \als(2)
    \,+\, 0.3248 \, \als(3) 
  \,+\, \ldots\, \right)
\; , \nn \\
  P_{\,\rm qq}^{}(8,4) &\,=\,& - 0.83224\als() \left(
  1 \,+\, 0.6328\, \als()
    \,+\, 0.4235\, \als(2)
    \,+\, 0.3121\, \als(3) 
  \,+\, \ldots\, \right)
\; , \\[4mm]
\label{Pqgnf4}
  P_{\,\rm qg}^{}(2,4) &\,=\,& \phantom{-}0.21221\, \als() \left(
  1 \,+\, 0.9004\, \als()
    \,-\, 0.1028\, \als(2)
    \,-\, 0.2367\, \als(3) 
    \,+\, \ldots\, \right)
\;, \nn \\
  P_{\,\rm qg}^{}(4,4) &\,=\,& \phantom{-}0.11671\, \als() \left(
  1 \,-\, 0.2801\, \als()
    \,-\, 0.9986\, \als(2)
    \,+\, 0.1297\, \als(3) 
    \,+\, \ldots\, \right)
\;, \nn \\
  P_{\,\rm qg}^{}(6,4) &\,=\,& \phantom{-}0.08337\, \als() \left(
  1 \,-\, 0.8389\, \als()
    \,-\, 1.1501\, \als(2)
    \,+\, 0.4417\, \als(3) 
    \,+\, \ldots\, \right)
\; , \\[4mm]
\label{Pgqnf4}
  P_{\,\rm gq}^{}(2,4) &\,=\,& \phantom{-}0.28294\, \als() \left(
  1 \,+\, 0.6219\, \als()
    \,+\, 0.1461\, \als(2)
    \,+\, 0.3622\, \als(3) 
    \,+\, \ldots\, \right)
\; , \nn \\
  P_{\,\rm gq}^{}(4,4) &\,=\,& \phantom{-}0.07781\, \als() \left(
  1 \,+\, 1.1152\, \als()
    \,+\, 0.8234\, \als(2)
    \,+\, 0.8833\, \als(3) 
    \,+\, \ldots\, \right)
\; , \nn \\
  P_{\,\rm gq}^{}(6,4) &\,=\,& \phantom{-}0.04446\, \als() \left(
  1 \,+\, 1.3019\, \als()
    \,+\, 1.0516\, \als(2)
    \,+\, 1.1270\, \als(3) 
    + \ldots\, \right)
\; , \nn \\
  P_{\,\rm gq}^{}(8,4) &\,=\,& \phantom{-}0.03116\, \als() \left(
  1 \,+\, 1.4309\, \als()
    \,+\, 1.1830\, \als(2)
    \,+\, 1.3184\, \als(3) 
    \,+\, \ldots\, \right)
\eea
and
\bea
\label{Pggnf4}
  P_{\,\rm gg}^{}(2,4) &\,=\,& -0.21221\, \als() \left(
  1 \,+\, 0.9004\, \als()
    \,-\, 0.1028\, \als(2)
    \,-\, 0.2367\, \als(3) 
    \,+\, \ldots\, \right)
\;, \nn \\
  P_{\,\rm gg}^{}(4,4) &\,=\,& -1.21489\, \als() \left(
  1 \,+\, 0.3835\, \als()
    \,+\, 0.1220\, \als(2)
    \,+\, 0.2405\, \als(3) 
    \,+\, \ldots\, \right)
\;, \nn \\
  P_{\,\rm gg}^{}(6,4) &\,=\,& -1.62755\, \als() \left(
  1 \,+\, 0.3937\, \als()
    \,+\, 0.1697\, \als(2)
    \,+\, 0.1902\, \als(3) 
    \,+\, \ldots\, \right)
\; . 
\eea
The corresponding analytic expressions for a general gauge group will 
be presented elsewhere.

\vspace*{1mm}
The relative size of the N$^2$LO and N$^3$LO contributions to 
eqs.~(\ref{Pqqnf4}) -- (\ref{Pggnf4}) is illustrated in fig.~1 for
$\al = 0.2$: The N$^3$LO corrections are less than 1\%, and less than
0.5\% of the NLO results except for $P_{\,\rm gq}^{}$, the quantity
with the lowest LO values, at $N \geq 4$.

\vspace{1mm}
The resulting low-$N$ expansion for the singlet evolutions equations
(\ref{qPsg}) is illustrated in fig.~2 for the sufficiently realistic
order-independent model input 
\bea  
\label{qsgInp}
  xq_{\sf s}^{}(x,\mu_{0}^{\:\!2}) & =  &
  0.6\: x^{\, -0.3} (1-x)^{3.5}\, \left(1 + 5.0\: x^{\, 0.8\,} \right)
\: , \nn \\[-0.5mm]
  x\:\!g (x,\mu_{0}^{\:\!2})\: & =  &
  1.6\: x^{\, -0.3} (1-x)^{4.5}\, \left(1 - 0.6\: x^{\, 0.3\,} \right)  
\eea
with $\al(\mu_{0}^{\:\!2}) = 0.2$ and $\nf=4$, which was already used 
in ref.~\cite{P2}.
The N$^3$LO corrections are very small at the standard renormalization 
scale $\mu_{r}^{}= \mu_{f}^{}\equiv \mu_{0}^{}$.
They lead to a reduction of the scale dependence to about 1\% 
(full width) at $N=4\,$ \& $N=6$ for the conventional range 
\mbox{$\frac{1}{4}\: \mu_{f}^{\:\!2} \leq \mu_{r}^{\:\!2} 
\leq 4\, \mu_{f}^{\:\!2}$}.

\vspace*{1mm}
Extending eqs.~(\ref{Pqgnf4}) and (\ref{Pggnf4}) to $N=8$ would be 
extremely hard with the hardware and software used to obtain these 
results; computing the $N=10$ results in this way is virtually
impossible.

\begin{figure}[thbp]
\centerline{\hspace*{-2mm}\epsfig{file=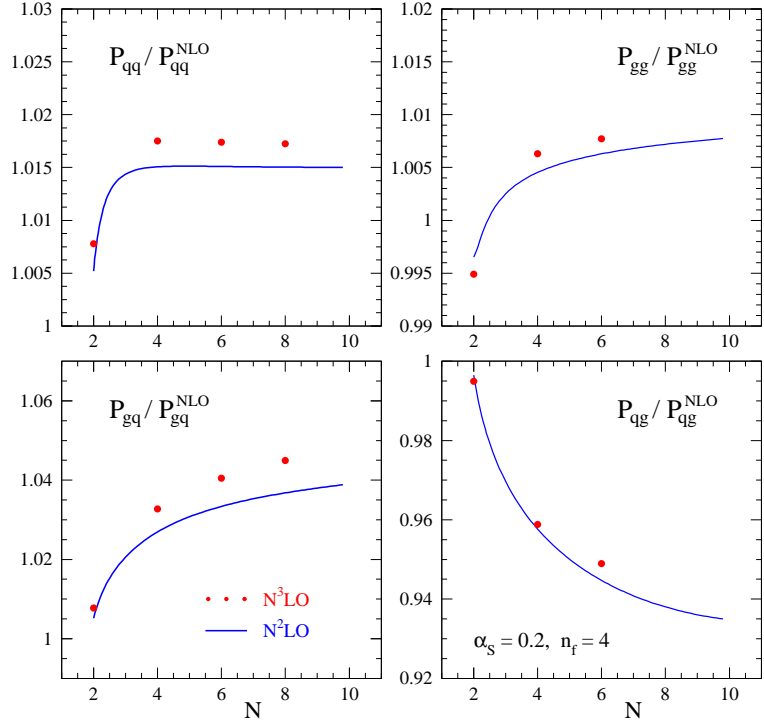,width=10cm,angle=0}}
\vspace{-2mm}
\caption{\label{Fig1}
Moments of the singlet splitting functions at NNLO (lines) and N$^3$LO
(even-$N$ points) for $\al = 0.2$ and $\nf = 4$, normalized to the 
respective NLO approximations.
}
\end{figure}
\begin{figure}[tbhp]
\centerline{\hspace*{-2mm}\epsfig{file=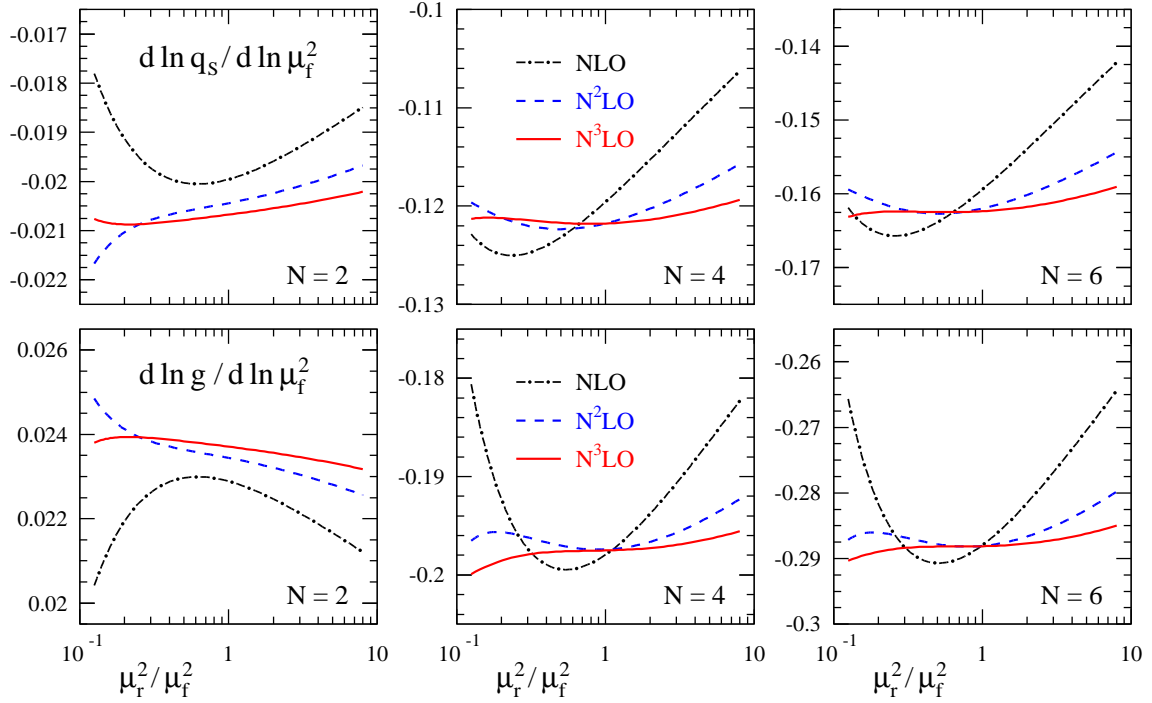,width=15cm,angle=0}}
\vspace{-2mm}
\caption{\label{Fig2}
The dependence of the logarithmic factorization-scale derivatives of 
the singlet PDFs on the renormalization scale $\mu_{r}^{}$ at $N=2$ 
(where the very small scaling violations of $q_{\sf s}^{}$ and $g$ are 
related by the momentum sum rule)
 $N=4$ and $N=6$ for the initial distributions (\protect\ref{qsgInp}). 
}
\end{figure}

\section{Quartic colour-factor contributions and the cusp anomalous dimensions}

\noindent
The computations of the four-loop splitting functions can be extended
to higher $N$ by using the OPE, since there the complexity of the
required self-energy integral increases by 2 for $N \ra N+2$ instead
of by 4 in the case of DIS.
For example, $N=16$ has been 
reached for the complete N$^3$LO contribution to $P_{\rm ns}^{\,+}$. 
In the limit of a large number of colours $\nc$, it was possible to 
reach \mbox{$N=20$}, which led to the determination of the all-$N$ 
expressions and hence of $P_{\rm ns}^{\,\pm (3) }(x)$ in this 
limit~\cite{MRUVV1}.

\vspace*{1mm}
In general the higher-order application of the OPE in massless
perturbative QCD is conceptually much more involved in the singlet 
case; for low-order treatments see refs.~\cite{OPEsgl}. This 
situation is far less severe for the contributions with quartic 
Casimir invariants,
\beq
\label{d4abbr}
  d^{\,(4)}_{xy} \;\equiv\; d_x^{\,abcd} d_y^{\,abcd}
\eeq
where $x,y$ labels the representations with generators $T_r^a$ and
\beq
\label{d4def}
  d_{r}^{\,abcd} \; =\; \frct{1}{6}\: {\rm Tr} \, ( \,
   T_{r}^{a\,} T_{r}^{b\,} T_{r}^{c\,} T_{r}^{d\,}
   + \,\mbox{ five $bcd$ permutations}\:\! ) 
\; ,
\eeq
which occur in the splitting functions for the first time at four 
loops. This effective `leading-order' situation implies particular 
relations and facilitates calculational simplifications. 
These include 
\beq
\label{SUSYrel}
    P_{\,\rm qq}^{\,(3)}(N) + P_{\,\rm gq}^{\,(3)}(N)
  - P_{\,\rm qg}^{\,(3)}(N) - P_{\,\rm gg}^{\,(3)}(N)
  \;\:\Qeq\:\: 0
\eeq
($\,\Qeq$ denotes equality for the quartic Casimir contributions)
for the colour-factor substitutions \cite{avLL16}
\beq
\label{SUSYcol}
      (2 \nf)^2 \, {d^{\,(4)}_{F\!F}}/{n_a}
 \;=\; 2 \nf \, {d^{\,(4)}_{FA}}/{n_a}
 \;=\; 2 \nf \, {d^{\,(4)}_{F\!F}}/{n_c}
 \;=\; {d^{\,(4)}_{FA}}/{n_c}
 \;=\; {d^{\,(4)}_{AA}}/{n_a}
\eeq
that lead to an ${\cal N}=1$ supersymmetric theory; for lower-order 
discussions see refs.~\cite{SUSY23loop}. Moreover the off-diagonal
quantities are found to be related by \cite{MRUVV2}
\beq
\label{Qoffd}
  P_{\,\rm qg}^{\,(0)}(N) \: P_{\,\rm gq}^{\,(3)}(N)
  \;\:\Qeq\:\:
  P_{\,\rm gq}^{\,(0)}(N) \: P_{\,\rm qg}^{\,(3)}(N)
\; .
\eeq
This second relation, which we have found empirically by inspecting
our results, is consistent with the implications of ${\cal N} = 1$
supersymmetry for QCD conformal operators investigated in 
ref.~\cite{BelitskyMS98}. It is also a special case of the structure
predicted in ref.~\cite{BassoK06} from the conformal symmetry of QCD 
at some non-integer space-time dimension $D = 4 - 2 \ep$. 

\vspace*{1mm}
We have used eqs.~(\ref{SUSYcol}) and (\ref{Qoffd}) partly to check
the results of our diagrams calculations, and partly to simplify our 
computational task at the highest values of $N$. 
In this manner, we have been able to derive all $d^{\,(4)}_{xy}$ 
contributions to the N$^3$LO splitting functions at $N \leq 16$. 
These results, and the structurally interesting all-$N$ expressions 
for the $\zr5$-terms, can be found in ref.~\cite{MRUVV2}. 

\vspace {1mm}
Analogous to the non-singlet quantities analyzed in ref.~\cite{MRUVV1},
the moments of $P_{\,\rm gg}^{\,(3)}$ at $N \leq 16$ facilitate 
numerical determinations of the quartic-Casimir contributions to the 
four-loop gluon cusp anomalous dimension $A_{4,\rm g}$, recall 
eq.~(\ref{xto1}). 
The present status of $A_{4,\rm q}$ and $A_{4,\rm g}$ is collected in 
table~1.

\vspace*{1mm}
The~coefficients of $A_{4,\rm q}$ which are known exactly have also 
been determined from the quark form factor \cite{FFnf2,qFF}; 
the results are in complete agreement. 
Recently, the exact coefficient of $\cft \nf$ has been obtained 
in ref.~\cite{Grozin18}.
The only piece of $A_{4,\rm g}$ known exactly so far is the $C_A \nft$ 
contribution \cite{DRUVV,gFFnf3}.
For numerical results in ${\cal N} \!=\! 4$ maximally supersymmetric 
Yang-Mills theory see ref.~\cite{BHY17}.  

\pagebreak

\begin{table}[th]
\vspace*{-1mm}
\centering
  \renewcommand{\arraystretch}{1.2}
  \begin{tabular}{MMMM}
   \mbox{quark}  & \mbox{gluon} &    A_{4,\rm q}             &  A_{4,\rm g}     \\[1pt]
   \hline\\[-5mm]
    \cff         &     -        &     0                      &        -              \\[1pt]
    \cft\, \ca   &     -        &     0                      &        -              \\[1pt]
    \cfs \cas    &     -        &     0                      &        -              \\[1pt]
    \cf \cat     &   \caf       & \phantom{-}  610.25 \pm 0.1~~ &                    \\[1pt]
    d^{\,(4)}_{F\!A}/N_{\!F}^{} & d^{\,(4)}_{A\!A}/N_{\!A}^{} 
                                &    -507.0 \pm 2.0          & -507.0 \pm 5.0~~      \\[0.5mm]
\hline\\[-5mm]
    \nf\, \cft   & \nf\,\cfs\ca &    -31.00554               &                          \\[1pt]
  \nf\,\cfs\ca   & \nf\,\cf\cas & \phantom{-} 38.75 \pm 0.2  &                          \\[1pt]
  \nf\,\cf\cas   & \nf\cat      &  -440.65 \pm 0.2~~         &                          \\[1pt]
    \nf\,d^{\,(4)}_{F\!F}/N_{\!F}^{} & \nf\,d^{\,(4)}_{F\!A}/N_{\!A}^{}   
                                &  -123.90  \pm 0.2~~        & -124.0 \pm 0.6~~         \\[0.5mm]
\hline\\[-5mm]
    \nfs\,\cfs   & \nfs\,\cf\ca &   -21.31439                &                          \\[-0.3mm]
  \nfs\,\cf\ca   & \nfs\,\cas   & \phantom{-}58.36737        &                          \\[-0.3mm]
       -         & \nfs\, d^{\,(4)}_{F\!F}/N_{\!A}^{}  
                                &      -                     & ~~\phantom{-}0.0 \pm 0.1 \\[-0.3mm]
    \nft\,\cf    & \nft\,\ca    & \phantom{-}2.454258        & ~~\phantom{-}2.454258    \\[0.2mm]
  \hline
  \end{tabular}
  \vspace*{2mm}
 \caption{\small \label{tab:AB}
 Fourth-order coefficients of the quark and gluon cusp anomalous 
 dimensions determined from the large-$x$ limit (\protect\ref{xto1}) 
 of the quark-quark and gluon-gluon splitting functions. 
 The errors in the quark case are correlated due to the exactly known 
 large-$n_c$ limit.
 The numerical value of $-31.00 \pm 0.4$ of ref.~\protect\cite{MRUVV1} 
 for the coefficient of $\nf \cft$ in $A_{4,\rm q}$ has been replaced 
 by the exact result of ref.~\protect\cite{Grozin18}. 
 This and the values for the $\nfs$ and $\nft$ coefficients have 
 been rounded to seven digits.
 Entries left blank for $A_{4,\rm g}$ have not been calculated from
 diagrams so far, but are related to those for $A_{4,\rm q}$ by
 Casimir scaling. Entries marked by `--' do not exist.
 }
 \vspace*{-1mm}
\end{table}

As up to the third order \cite{P2},
the corresponding quark and gluon entries in table 1 have the same 
coefficients (for now: as far as they have been computed, and within 
numerical errors).  We refer to this (for now: conjectured) relation 
as {\em generalized Casimir scaling}.

\vspace*{1mm}
Unlike for the lower-order coefficients, this relation does not have 
the consequence that 
the values of $A_{4,\rm g}$ and $A_{4,\rm q}$ are related by a simple 
numerical Casimir scaling in QCD, i.e., a factor of $C_A / C_F = 9/4$.
However, this numerical Casimir scaling is restored in the large-$n_c$ 
limit of the quartic colour factors, and therefore also in the overall 
large-$n_c$ limit, see also ref.~\cite{Dixon17}.

\vspace*{1mm}
The results in table 1 and the generalized Casimir scaling lead to the
following numerical results for the four-loop cusp anomalous dimensions
in QCD, expanded in powers of $\al/(4\pi)$:
\bea
\label{A4q}
  A_{4,\rm q} &\,=\,&   20702(2)\phantom{0} \,-\, 5171.9(2) \,\nf
                  \,+\, 195.5772\,\nfs      \,+\, 3.272344 \,\nft
\; , \\[-0.5mm]
\label{A4g}
  A_{4,\rm g} &\,=\,&   40880(30)      \,-\, 11714(2) \;\nf\,
                  \,+\, 440.0488\,\nfs \,+\, 7.362774 \,\nft
\; ,
\eea
where the number(s) in brackets indicate the uncertainty of the
preceding digit(s).
Combining these results with the lower-order coefficients,
one arrives at the very benign expansions
\bea
\label{Aqnf345}
 A_{\rm q}(\al,\nf\!=\!3) &=& 0.42441\,\al \,
 [\, 1 +  0.72657\, \al +  0.73405\, \als(2) + 0.6647(2)\, \als(3) + \ldots\, ]
\; , \nn \\[-0.2mm]
 A_{\rm q}(\al,\nf\!=\!4) &=& 0.42441\,\al \,
 [\, 1 +  0.63815\, \al +  0.50998\, \als(2) + 0.3168(2)\, \als(3) + \ldots\, ]
\; , \nn \\[-0.2mm]
 A_{\rm q}(\al,\nf\!=\!5) &=& 0.42441\,\al \,
 [\, 1 +  0.54973\, \al +  0.28403\, \als(2) + 0.0133(3)\, \als(3) + \ldots\, ]
\eea
and
\bea
\label{Agnf345}
  A_{\rm g}(\al,\nf\!=\!3) &=& 0.95493\,\al \,
  [\, 1 + 0.72657\, \al  +  0.73405\, \als(2) + 0.415(2)\, \als(3) + \ldots\, ]
\; , \nn \\[-0.2mm]
  A_{\rm g}(\al,\nf\!=\!4) &=& 0.95493\,\al \,
  [\, 1 + 0.63815\, \al  +  0.50998\, \als(2) + 0.064(2)\, \als(3) + \ldots\, ]
\; , \nn \\[-0.2mm]
  A_{\rm g}(\al,\nf\!=\!5) &=& 0.95493\,\al \,
  [\, 1 + 0.54973\, \al  +  0.28403\, \als(2) - 0.243(2)\, \als(3) + \ldots\, ]
\; . 
\eea
The remaining uncertainties of the N$^3$LO coefficients 
are practically irrelevant for phenomenological applications. 
Note that,
due to the breaking of the numerical Casimir scaling especially 
in the $\nfz$ parts of eqs.~(\ref{A4q}) and (\ref{A4g}) and the 
cancellations between the terms without and with $\nf$, 
the~numerical Casimir scaling is completely broken for the N$^3$LO 
terms in eqs.~(\ref{Aqnf345}) and (\ref{Agnf345}).

\section{First results for the N$\bm ^4$LO non-singlet splitting functions}

\noindent
Using the recent implementation \cite{RstarN} of the local 
R$^\ast$-operation \cite{RstarC}, it is now possible, at least for 
the lowest values of $N$, to extend the {\sc Forcer} calculations 
of the splitting functions to the N$^4$LO contributions $P^{\,(4)}$ 
in eq.~(\ref{Pexp}).
The computational setup is similar to (but includes some efficiency
improvements upon) that used for the beta function and Higgs decays 
at five loops in refs.~\cite{5loopN1,5loopN2}.

\vspace*{1mm}
As a check specific to the present case, we have explicitly verified
that $P_{\rm ns}^{\,(4)-}(N\!=\!1)$ vanishes in a calculation with 
one power of the gauge parameter. We have then calculated 
$P_{\rm ns}^{\,(4)+}(N=2)$ and $P_{\rm ns}^{\,(4)-}(N=3)$ for a 
general gauge group. The latter computation required an effort 
comparable to that for the N$^4$LO corrections to $H \ra gg$ in the
heavy top-quark limit refs.~\cite{5loopN2}, the hardest calculation
performed so far with the program of ref.~\cite{RstarN}.  
An extension to $N=4$ would be extremely hard with the present tools;
higher values of $N$ are out of reach for now.

\vspace*{1mm}
The analytic results will be presented elsewhere. Before 
turning to their numerical effects, it is worthwhile to mention
another, if not particularly strong check: besides rational numbers,
the moments of the N$^4$LO splitting functions include values of 
Riemann's $\zeta$-function up to $\zr7$. 
Consistent with the `no-$\pi^2\!$ 
theorem' for Euclidean physical quantities \cite{no-pi2}, the $\zr6$ 
terms disappear when the \MSb\ splitting functions are converted to
physical evolution kernels for structure functions in DIS, and the 
$\zr4$ terms disappear after transforming to a renormalization 
scheme in which the N$^4$LO beta function does not include 
$\zr4$-terms, such as {\sc MiniMOM} in the Landau gauge
\cite{MiniMOM,4LoopPrp}.

\vspace*{1mm}
Our new results for $N=2$ and $N=3$ lead to the numerical \MSb\
expansions  
\bea
\label{PnspN2}
  P_{\rm ns}^{+}(2,0) &=&
  - 0.2829\:\! \als() ( 1 \,+\, 1.0187\:\! \als() \,+\, 1.5307\:\! \als(2)
  + 2.3617\:\! \als(3) {  \,+\, 4.520 \:\!\als(4) } + \,\ldots\, )
\; , \nn \\[-1mm] &\cdots& \nn \\[-1mm]
  P_{\rm ns}^{+}(2,3) &=&
  - 0.2829 \:\!\als() ( 1 \,+\, 0.8695\:\! \als() \,+\, 0.7980\:\! \als(2)
  + 0.9258\:\! \als(3) {  \,+\, 1.781 \:\!\als(4) } + \,\ldots\, )
\; , \nn \\
  P_{\rm ns}^{+}(2,4) &=&
  - 0.2829\:\! \als() ( 1 \,+\, 0.7987\:\! \als() \,+\, 0.5451\:\! \als(2)
  + 0.5215\:\! \als(3) {  \,+\, 1.223 \:\!\als(4) } + \,\ldots\, )
\; , \nn \\
  P_{\rm ns}^{+}(2,5) &=&
  - 0.2829\:\! \als() ( 1 \,+\, 0.7280\:\! \als() \,+\, 0.2877\:\! \als(2)
  + 0.1512\:\! \als(3) {  \,+\, 0.849 \:\!\als(4) } + \,\ldots\, )
\eea
and
\bea
\label{PnsmN3}
  P_{\rm ns}^{-}(3,0) &=&
  - 0.4421\:\! \als() ( 1 \,+\, 1.0153\:\! \als() \,+\, 1.4190\:\! \als(2)
  + 2.0954\:\! \als(3) {  \,+\, 3.954 \:\!\als(4) } + \,\ldots\, )
\; , \nn \\[-1mm] &\cdots& \nn \\[-1mm]
  P_{\rm ns}^{-}(3,3) &=&
  - 0.4421 \:\!\als() ( 1 \,+\, 0.7952\:\! \als() \,+\, 0.7183 \als(2)
  + 0.7605\:\! \als(3) {  \:+\, 1.508 \:\!\als(4) } + \,\ldots\, )
\; , \nn \\
  P_{\rm ns}^{-}(3,4) &=&
  - 0.4421 \:\!\als() ( 1 \,+\, 0.7218\:\! \als() \,+\, 0.4767\:\! \als(2)
  + 0.3921\:\! \als(3) {  \,+\, 1.031 \:\!\als(4) } + \,\ldots\, )
\; , \nn \\
  P_{\rm ns}^{-}(3,5) &=&
  - 0.4421 \:\!\als() ( 1 \,+\, 0.6484\:\! \als() \,+\, 0.2310\:\! \als(2)
  + 0.0645\:\! \als(3) {  \,+\, 0.727 \:\!\als(4) } + \,\ldots\, )
\; .
\eea
Here we have included $\nf = 0$ besides the physically relevant values,
since it provides useful information about the behaviour of the
perturbation series. The N$^4$LO coefficients in (\ref{PnspN2}) and
(\ref{PnsmN3}) are larger than one may have expected from the NNLO 
and N$^3$LO contributions. 

\vspace*{1mm}
It is interesting in this context to consider the effect of the quartic 
group invariants.
For example, the $\nf = 0$ coefficients in eqs.~(\ref{PnspN2}) and 
(\ref{PnsmN3}) at N$^3$LO and N$^4$LO can be decomposed as
\bea
  2.3617 &=& 2.0878 \,+\, 0.1096\, d^{\,(4)}_{FA}/{n_c}
\nn \\
  4.520\phantom{0} 
         &=& 3.552\phantom{1} 
                    \,-\, 0.0430\, d^{\,(4)}_{FA}/{n_c}
                    \,+\, 0.0510\, d^{\,(4)}_{AA}/{n_a}
\eea
and
\bea
  2.0954 &=& 2.0624 \,+\, 0.0132\, d^{\,(4)}_{FA}/{n_c}
\nn \\
  3.954\phantom{1}
         &=& 3.371\phantom{1}
                    \,-\, 0.0171\, d^{\,(4)}_{FA}/{n_c}
                    \,+\, 0.0371\, d^{\,(4)}_{AA}/{n_a}
\eea
with $d^{\,(4)}_{FA}/{n_c} = 5/2$ and $d^{\,(4)}_{AA}/{n_a} = 135/8$
in QCD, see, e.g., app.~C of ref.~\cite{4LoopPrp}: Without the 
rather large contributions of $d^{\,(4)}_{AA}$, which enter at  
N$^4$LO for the first time, the series would look much more benign 
with consecutive ratios of 1.4 to 1.6 between the N$^4$LO, N$^3$LO, 
NNLO and NLO coefficients.
This sizeably $d^{\,(4)}_{AA}$ contribution ($\,\sim \ncs + 36\,$) 
also implies that the leading large-$\nc$ contribution provides a 
less good approximation at N$^4$LO than at the previous orders.  

\vspace*{1mm}
The numerical impact of the higher-order contributions to the 
splitting functions $P_{\rm ns}^{\,\pm}$ on the $N=2$ and $N=3$
moments of the respective PDFs (\ref{qns}) are illustrated in fig.~3.
At $\al(\mu_{f}^{\:\!2}) = 0.2$ and $\nf=4$, the N$^4$LO corrections
are about 0.15\% at $\mu_{r}^{} = \mu_{f}^{}$, roughly half the size 
of their N$^3$LO counterparts. Varying $\mu_{r}^{}$ up and down by 
a factor of 2 ~--~ the required additional terms for the splitting
functions can be found to N$^4$LO, e.g., in eq.~(2.9) of 
ref.~\cite{NV4} ~--~ one arrives at a band with a full width of about
0.7\%. The N$^3$LO and N$^4$LO corrections are about twice as large
at a lower scale with $\al(\mu_{f}^{\:\!2}) = 0.25$ and $\nf=3$.

\begin{figure}[hbt]
\vspace{2mm}
\centerline{\hspace*{-2mm}\epsfig{file=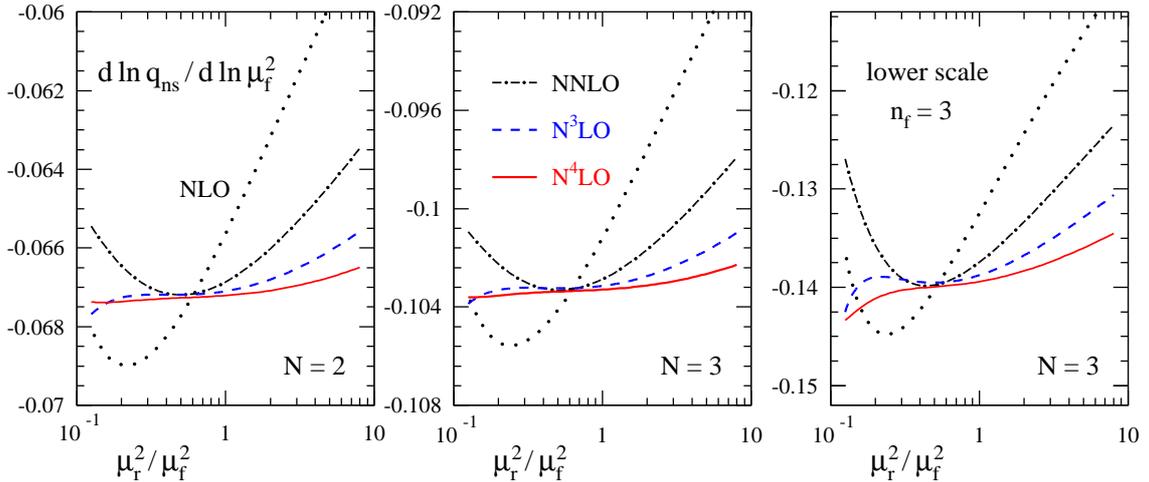,width=15cm,angle=0}}
\vspace{-2mm}
\caption{\label{Fig3}
Left and middle panel: the renormalization-scale dependence of 
the logarithmic factorization-scale derivatives of the PDFs 
$q_{\rm ns}^{\,+}$ at $N=2$ and $q_{\rm ns}^{\,-}$ at $N=3$ at our
standard reference point with $\al(\mu_{f}^{\:\!2}) = 0.2$ and $\nf=4$.
Right panel: the corresponding $N=3$ results at a lower scale with
$\al(\mu_{f}^{\:\!2}) = 0.25$ and $\nf=3$.  
}
\vspace*{-2mm}
\end{figure}

\hspace*{\fill} \newpage

\subsection*{Acknowledgements}

\noindent
A.V. is grateful to the organizers for support that facilitated his 
participation in this workshop.
The research reported here has been supported by the Advanced Grant 
320651, {\it HEPGAME}, of the {\it European Research Council}$\,$ (ERC) 
and by the {\it Deutsche Forschungsgemeinschaft} (DFG) under grant 
number MO~1801/2-1.

\end{document}